\documentclass[aps,pra,showpacs,amsmath,amssymb,amsfonts,floatfix,superscriptaddress,twocolumn,lengthcheck]{revtex4-1}
\usepackage{graphicx}
\usepackage{subfigure}
\usepackage{verbatim}
\usepackage{dcolumn}
\usepackage{bm}
\usepackage{epsf}
\usepackage{xcolor}
\usepackage{hyperref}
\usepackage{hhline}
\usepackage{float}
\usepackage{enumerate}
\usepackage{bbm}
\usepackage{lipsum}
\definecolor{mygreen}{rgb}{0.01, 0.31, 0.59}
\definecolor{myblue}{rgb}{0.01, 0.31, 0.59}
\hypersetup{
	colorlinks=true,   
	hypertexnames=false,
	linkcolor=blue,  
	citecolor=blue, 
	urlcolor =blue    
}

\begin{document}
	
	\title{Observation of magnon cross-Kerr effect in cavity magnonics}
	
	\author{Wei-Jiang Wu}
	\affiliation{Interdisciplinary Center of Quantum Information and Zhejiang Province Key Laboratory of Quantum Technology and Device, Department of Physics and State Key Laboratory of Modern Optical Instrumentation, Zhejiang University, Hangzhou 310027, China}
	
	\author{Da Xu}
	\email{daxu@zju.edu.cn}
	\affiliation{Interdisciplinary Center of Quantum Information and Zhejiang Province Key Laboratory of Quantum Technology and Device, Department of Physics and State Key Laboratory of Modern Optical Instrumentation, Zhejiang University, Hangzhou 310027, China}
	\email{daxu@zju.edu.cn}
	
	\author{Jie Qian}
	\affiliation{Interdisciplinary Center of Quantum Information and Zhejiang Province Key Laboratory of Quantum Technology and Device, Department of Physics and State Key Laboratory of Modern Optical Instrumentation, Zhejiang University, Hangzhou 310027, China}
	
	\author{Jie Li}
	\affiliation{Interdisciplinary Center of Quantum Information and Zhejiang Province Key Laboratory of Quantum Technology and Device, Department of Physics and State Key Laboratory of Modern Optical Instrumentation, Zhejiang University, Hangzhou 310027, China}

	\author{Yi-Pu Wang}
	\email{yipuwang@zju.edu.cn}
	\affiliation{Interdisciplinary Center of Quantum Information and Zhejiang Province Key Laboratory of Quantum Technology and Device, Department of Physics and State Key Laboratory of Modern Optical Instrumentation, Zhejiang University, Hangzhou 310027, China}
	
	\author{J. Q. You}
	\email{jqyou@zju.edu.cn}
	\affiliation{Interdisciplinary Center of Quantum Information and Zhejiang Province Key Laboratory of Quantum Technology and Device, Department of Physics and State Key Laboratory of Modern Optical Instrumentation, Zhejiang University, Hangzhou 310027, China}
	
	\date{\today}
	
	\begin{abstract}
		When there is a certain amount of field inhomogeneity, the biased ferrimagnetic crystal will exhibit the higher-order magnetostatic (HMS) mode in addition to the uniform-precession Kittel mode. In cavity magnonics, we show both experimentally and theoretically the cross-Kerr-type interaction between the Kittel mode and HMS mode. When the Kittel mode is driven to generate a certain number of excitations, the HMS mode displays a corresponding frequency shift, and vice versa. The cross-Kerr effect is caused by an exchange interaction between these two spin-wave modes. Utilizing the cross-Kerr effect, we realize and integrate a multi-mode cavity magnonic system with only one yttrium iron garnet (YIG) sphere. Our results will bring new methods to magnetization dynamics studies and pave a way for novel cavity magnonic devices by including the magnetostatic mode-mode interaction as an operational degree of freedom.
		
	\end{abstract}
	
	\maketitle
	
	
	\section{INTRODUCTION}
	
	Nonlinear effects are a large class of physical phenomena that have been continuously and actively explored in optics~\cite{Boyd,Autere2018,Zhang2019}, plasmonics~\cite{Kauranen2012,Mesch2016,Zhong2020}, and acoustics~\cite{Hamilton1998,Fang2017,Darabi2019}, as well as in other oscillations and wave systems~\cite{Nayfeh2008,Kartashov2019,Zuo2020}. Nonlinearity and nonlinear interactions are weak and difficult to be detected in some systems, but they can also become sufficiently strong to be the dominant factor in some other systems. The cross-Kerr effect is one of the subtle nonlinear interactions between fields and waves that can exist in natural ions~\cite{Ding2017}, atoms~\cite{He2014,Xia2018,Sinclair2019}, and artificial mesoscopic architectures, such as superconducting circuits~\cite{Hoi2013,Kounalakis2018,Vrajitoarea2020}. The cross-Kerr effect is a nonlinear shift in the frequency of a resonator as a function of the number of excitations in another mode that interacts with the resonator. Understanding these nonlinear interactions is not only of fundamental importance, but also useful in various applications. In practice, the cross-Kerr effect can be utilized to implement quantum logic gates~\cite{Turchette1995,Brod2016}, prepare entangled photons~\cite{Lukin2000,Sheng2008}, and carry out quantum non-demolition measurements~\cite{Wang2019,Dassonneville2020}.
	
	Cavity magnonics, on the other hand, has gradually demonstrated its unique advantages in fundamental and applied research over the last few years~\cite{Huebl2013,Tabuchi2014,Zhang2014,Goryachev2014,Bai2015,Cao2015,Zhang2015,Tabuchi2015,Haigh2016,Maier-Flaig2016,A.Osada2016,X.Zhang2018,Zhang2017,PuWang2018,A.Osada2018,JieLi2018,Harder2018,Y.P.Wang2019,Lachance-Quirion2020,Y.Yang2020,J.Xu2020,S.P.Wolski2020,Jinwei2021,Xufeng2021,Potts2021,D.Zhang2015}. It is expected to be a critical component of hybrid quantum systems~\cite{LachanceQuirion2019} and quantum network nodes~\cite{Jie2021}. The commonly used physical implementation of a cavity magnonic system consists of a microwave cavity and \textit{ferro-}(\textit{ferri-})magnets, such as YIG. The spin collective excitation modes are spin waves in the ferrimagnetic spin ensemble. The most widely studied spin wave mode is the uniform precession mode of spins (i.e., the Kittel mode), which has the largest magnetic dipole moment and can strongly interact with the microwave in both the quantum limit~\cite{Huebl2013,Tabuchi2014} and room-temperature classical regime~\cite{Zhang2014}. Under the bias of an external magnetic field with certain inhomogeneity, besides the Kittel mode, there are also higher-order magnetostatic (HMS) modes in the spin ensemble~\cite{J.F.Dillon1958,P.C.Fletcher1959}. Previous studies often ignored the influence of these magnetostatic modes on the observation and application of Kittel mode in experiments. However, the HMS modes actually have nontrivial spin textures. Their non-zero orbital angular momentum feature can be used to demonstrate nonreciprocal Brillouin light scattering in optomagnonics and other chiral optics~\cite{A.Osada2018,J.Graf2018,J.A.Haigh2018,Osada_2018}. In the case of YIG sphere supporting optical whispering-gallery modes, the optomagnonic coupling strength can be enhanced for HMS modes due to their reduced mode volume and the localized distribution on the boundary~\cite{J.Graf2018,A.Gloppe2019}, which increases the microwave-to-optical photon transduction efficiency in the cavity optomagnonic system.
	
	In this work, we focus on the cavity magnonic system with the cross-Kerr interaction between the Kittel mode and HMS mode, used as a new control degree of freedom. This cross-Kerr interaction is investigated both theoretically and experimentally in a coupled 3D cavity and YIG sphere system, where the YIG sphere supports both the Kittel mode and HMS mode. A large frequency detuning between these two modes is found. When the Kittel mode or HMS mode is pumped to generate a certain number of magnon excitations, the pumped mode gains a frequency shift due to the self-Kerr effect~\cite{PuWang2018,Y.P.Wang2016,GQ2019,Kong2019,Xiao2021,Ruichang2021,Gurevich2020}. Meanwhile, the frequency of the undriven mode also shifts due to the cross-Kerr effect between these two modes. Then, the ratio of the self-Kerr and cross-Kerr coefficients can be obtained. Furthermore, using the cross-Kerr coefficient as a ruler, we find that the magnetostatic mode's self-Kerr coefficient is greater than that of the Kittel mode. This offers additional controllable degrees of freedom in cavity magnonics without increasing system's complexity. Our study will also bring new ideas to the fundamental magnetostatics studies and hybrid magnonic operations~\cite{Y.Li2020,B.Z.R,Yuan}.
	
	\begin{figure}[!t]
		\centering
		\includegraphics[width=0.45\textwidth]{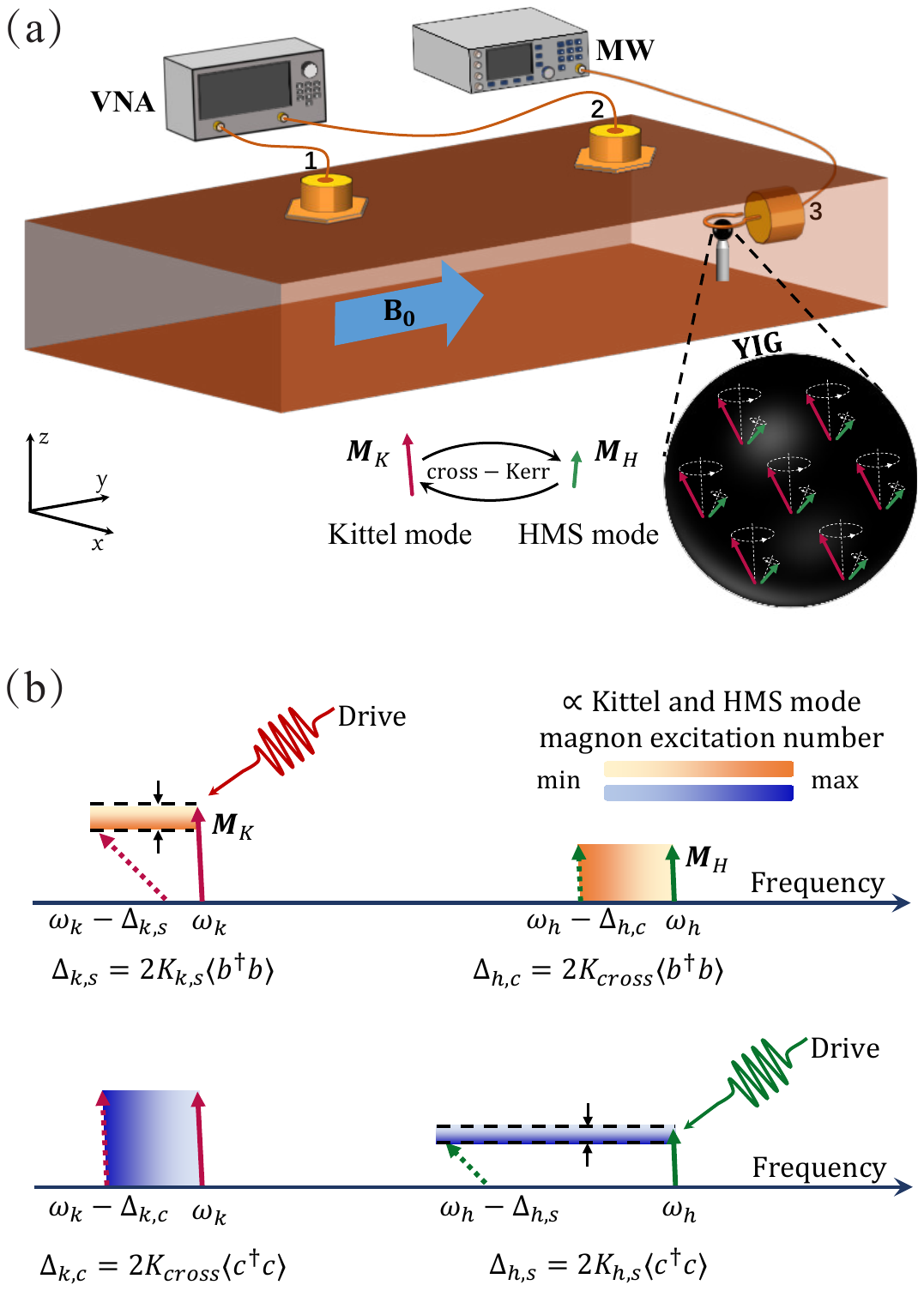}
		\caption{(a) Schematic diagram of the experimental setup. The YIG sphere is coupled to the cavity mode $TE_{102}$. The cavity is placed in the bias magnetic field $B_0$ (blue arrow). The microwave source connected to the loop antenna provides a drive field perpendicular to the bias magnetic field. The vector network analyzer (VNA) measures the cavity transmission spectrum $|S_{21}|^2 $. Inset: in the YIG sphere, the Kittel mode and HMS mode are denoted by two sub-magnetizations. (b) Schematic diagram of the self-Kerr and cross-Kerr effect. In the upper panel, the drive field applied on the Kittel mode will excite a certain number of magnons and enlarge the precession angle of the Kittel-mode magnetization $M_K$. The self-Kerr effect will cause a frequency shift of the Kittel mode. The cross-Kerr effect will result in a frequency shift of the HMS mode without changing the precession angle of the HMS-mode magnetization $M_H$. A similar effect occurs by exchanging the roles of the two modes, as shown in the lower panel.}
		\label{Figure1}
	\end{figure}

	\section{SYSTEM AND THEORETICAL MODEL}
	The experimental setup of the cavity magnonic system is illustrated in Fig.~\ref{Figure1}(a). It consists of the microwave field in a 3D copper cavity (internal dimension is 44.0$\times$20.0$\times $6.0 mm$^3$ ), the magnon modes in a single crystal YIG sphere (1 mm in diameter), and the drive field provided by a microwave (MW) source. The YIG sphere is glued on the top of a beryllium ceramic rod and mounted at the antinode of the magnetic field of the cavity mode $TE_{102} $ ($ {\omega _c}/2\pi = 10.07$ GHz). The whole sample is placed in a bias magnetic field $B_0$ applied along the $ \left[{110}\right] $ crystal axis, which results in a negative self-Kerr coefficient~\cite{PuWang2018,Gurevich2020}. The bias magnetic field, the magnetic field of the cavity mode $TE_{102} $, and the drive field magnetic component are perpendicular to each other at the position of the YIG sphere. This configuration maximizes the cavity-YIG coupling strength and drive efficiency. The drive field is loaded to port 3, which is terminated with a loop antenna. The vector network analyzer (VNA) is used for measuring the microwave transmission spectrum $|S_{21}|^2$ via port 1 and port 2. The probe power is much weaker than the drive power. It should be noted that the loop antenna plays multiple roles: (i) radiates the drive field; (ii) works as a dissipative channel for the magnon modes; (iii) creates an inhomogeneous magnetic field, which gives rise to the emergence of HMS modes.
	
	\begin{figure*}[!t]
		\centering
		\includegraphics[width=0.95\textwidth]{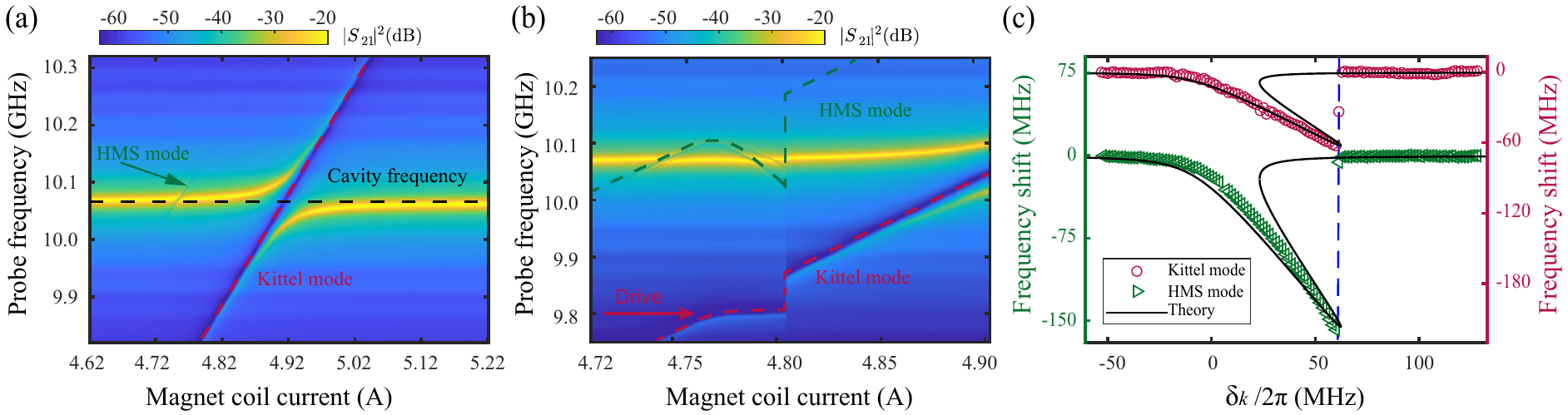}
		\caption{Drive on the Kittel mode. (a) The transmission spectra are measured versus magnet coil current. The Kittel mode (red dashed line) and HMS mode (green dashed line) are coupled with the cavity mode. (b) The transmission spectra are measured versus magnet coil current, while a drive field of 9.8 GHz and 25 dBm is applied on the Kittel mode. The Kittel mode has a frequency shift $\Delta_{k,s}$ due to the self-Kerr effect, and the HMS mode has a even larger frequency shift $\Delta_{h,c}$ caused by the cross-Kerr effect. (c) The frequency shifts of the Kittel mode and HMS mode are extracted from (b), and plotted versus $\delta_{k} = \omega_{k} -\omega_d $. The experimental results are fitted by Eq.~(\ref{eqution2}) and ${\Delta _{h,c}}=\frac{K_{cross}}{K_{k,s}}\Delta_{k,s}$, which also gives the ratio $K_{cross}/K_{k,s}=2.5$. }
		\label{Figure2}
	\end{figure*}

	Owing to the inhomogeneity of the magnetic field, apart from the Kittel mode, the HMS mode is also excited. The two modes are both supported by the collective spin motion. More spin moments contribute to the spin uniform precession mode (Kittel mode), yielding a large dipole moment for the Kittel mode. Therefore, the couplings between the Kittel mode and other electromagnetic fields are usually strong. However, fewer spin moments contribute to the dipole of the HMS mode, so the coupling between the HMS mode and cavity mode is weak in cavity magnonics~\cite{LachanceQuirion2019,C.Zhang2021}. Both the Kittel mode and HMS mode are nonlinear modes on account of the magnetocrystalline anisotropy. The magnetocrystalline anisotropy energy measures the energy cost when tuning the magnetization orientation from the easy axis to the hard axis. Mathematically, the energy term is proportional to $S_{z}^2$ ($S_{z}$ is the macrospin operator) and a self-Kerr term $K_{s}m^\dag mm^\dag m$ can be obtained, where $K_s$ is the Kerr coefficient and $m^\dag$ ($m$) is the creation (annihilation) operator of the magnon mode~\cite{Y.P.Wang2016,GQ2019}. When the magnon mode is pumped to generate a certain number of excitations, the precession angle of the magnetization increases. This process leads to the change of the magnetocrystalline anisotropy energy, which results in the frequency shift of the nonlinear mode. In previous studies, the interaction between the Kittel mode and HMS mode has rarely been investigated. Here, we model the two magnon modes with two sub-magnetizations ($M_{K}$ and $M_{H}$), which nonlinearly interact with each other. The interaction term is proportional to $S_{K}\cdot S_{H}$ (subscript $K$ and $H$ represent the Kittel mode and HMS mode), which gives a two-mode cross-Kerr term. The total Hamiltonian of the coupled system is (see Appendix A)
	\begin{equation}
		\begin{split}
			H/\hbar&= \omega _{c}a^\dag a + \omega _{k}b^\dag b+ \omega_{h}c^\dag c\\
			&+K_{k,s}b^\dag bb^\dag b + K_{h,s}c^\dag cc^\dag c+K_{cross}b^\dag bc^\dag c \\
			&+g_{k}\left(a^\dag b + b^\dag a \right)+g_{h}\left( a^\dag c + c^\dag a \right)+g_{k,h}\left( b^\dag c + c^\dag b \right)\\
			&+\Omega _k(b^\dag e^{ - i{\omega _d}t} + b e^{i{\omega _d}t})+\Omega _h(c^\dag e^{ - i{\omega _d}t} + c e^{i{\omega _d}t})
			\label{equation1}
		\end{split}
	\end{equation}
	where $a^\dag$ ($a$), $b^\dag$ ($b$) and $c^\dag$ ($c$) are the creation (annihilation) operators of the cavity mode at frequency $\omega_c$, the Kittel mode at frequency $\omega_{k}$, and the HMS mode at frequency $\omega_{h}$, respectively; $K_{k,s}$ and $K_{h,s}$ are the self-Kerr coefficients of the Kittel mode and HMS mode; $K_{cross}$ is the cross-Kerr coefficient between the two magnon modes; $g_{k}$ ($g_{h}$) is the coupling strength between the Kittel (HMS) mode and cavity mode; $g_{k,h}$ is the coherent coupling strength between the Kittel mode and HMS mode (see Appendix A); ${\Omega_k}$ (${\Omega_h}$) is the drive-field strength on the Kittel (HMS) mode, and $\omega_d$ is the drive-field frequency.

	The observable effects of the self-Kerr and cross-Kerr effects are schematically illustrated in Fig.\ref{Figure1}(b). The drive field applied on the Kittel mode enlarges the precession angle of the magnetization $M_{K}$, which also corresponds to the generation of a certain number of magnon excitations. In the case of Kittel mode with a negative self-Kerr coefficient and under the mean-field approximation, the frequency of the Kittel mode will have a red-shift $\Delta_{k,s}$~\cite{PuWang2018,Ruichang2021}, equals to $2K_{k,s}\langle b^{\dagger}b\rangle$. Here the extra subscript $s$ represent the self-Kerr. Simultaneously, the cross-Kerr effect will also induce a frequency shift of the undriven HMS mode, and the precession angle of the magnetization $M_{H}$ is, however, unchanged. The frequency shift caused by the cross-Kerr effect is also proportional to the Kittel-mode magnon excitation number, which is $\Delta_{h,c}=2K_{cross}\langle b^{\dagger}b\rangle$. Here the latter subscript $c$ represents the cross-Kerr. A similar mechanism occurs when the drive field is applied solely on the HMS mode, as shown in the bottom panel of Fig.~\ref{Figure1}(b). It is worth mentioning that the scale ratios of the frequency shifts drawn in the figure are not set arbitrarily but roughly correspond to the experimental results discussed in the following section.

	\section{EXPERIMENTAL RESULTS}
	
	The transmission spectra of the coupled system are measured versus the magnet coil current as shown in Fig.~\ref{Figure2}(a). The large anti-crossing indicates the strong coupling between the Kittel mode and cavity mode, and the coherent coupling strength is 40.5 MHz. At the left part of the transmission mapping, when the magnet coil current is about 4.75 A, we find a small coupling feature, which is attributed to the HMS mode. The coupling strength is 2 MHz. The detuning between the Kittel mode and HMS mode is 301 MHz.
	
	\subsection{Drive the Kittel mode}
	In Fig.~\ref{Figure2}(b), we apply the drive field with a power of 25 dBm at the fixed frequency of 9.8 GHz on the Kittel mode and measure the transmission spectra versus the magnetic coil current.  Typical bistable frequency jumps can be observed in both the Kittel mode and HMS mode, induced by the self-Kerr effect and cross-Kerr effect, respectively. This measurement is equivalent to sweeping the frequency of the drive field while fixing the frequencies of the magnon modes. We define the drive field detuning as $\delta_{k}=\omega_{k}-\omega_{d}$. It should be noted that our system works in the double-dispersive regime, where $\Lambda_{c,k}=\omega_{c}-\omega_{k}\gg g_{k}$ and $\Lambda_{h,k}=\omega_{h}-\omega_{k}\gg g_{k,h}$. The coherent couplings induced frequency shifts approximately equal $g_{k}^2/\Lambda_{c,k}$ and $g_{k,h}^2/\Lambda_{h,k}$, respectively~\cite{Y.P.Wang2016,GQ2019}. In the dispersive regime, they are all small terms. Also, the coupling strength between the HMS mode and cavity mode is relatively small.  In this case, we can intently pay attention to the Kerr nonlinearity induced effects. Then we extract and plot the frequency shifts of the two magnon modes versus $\delta_{k}$ in Fig.~\ref{Figure2}(c). The driven Kittel mode has a maximum frequency shift of about -60 MHz, and the undriven HMS mode has a synchronous frequency shift of about -150 MHz.
	
	The frequency shift $\Delta _{k,s}$ of the driven Kittel mode can be fitted by the following equation~\cite{Y.P.Wang2016,GQ2019}:
	\begin{equation}
		\left[ {{{\left( {{\Delta _{k,s}} + {\delta _k}} \right)}^2} + {(\gamma _k/2)}^2} \right]{\Delta _{k,s}} - {c_k}{P_d} = 0,
		\label{eqution2}
	\end{equation}
	where ${\gamma _{k}/2\pi}=11.6$ MHz is the Kittel mode linewidth, $c_k$ is the coefficient of drive efficiency on the Kittel mode, and $P_d$ is the drive power. For the undriven HMS mode, the frequency shift can be fitted by ${\Delta _{h,c}}=\frac{K_{cross}}{K_{k,s}}\Delta_{k,s}$. The fitting curves are shown as solid black lines in Fig.~\ref{Figure2}(c), which are in good agreement with the experimental results. We get the ratio of $K_{cross}/K_{s,k}=$2.5, which indicates that the cross-Kerr nonlinearity is 2.5 times larger than the self-Kerr nonlinearity of the Kittel mode.
	
	\begin{figure}[!htbp]
		\centering
		\includegraphics[width=0.45\textwidth]{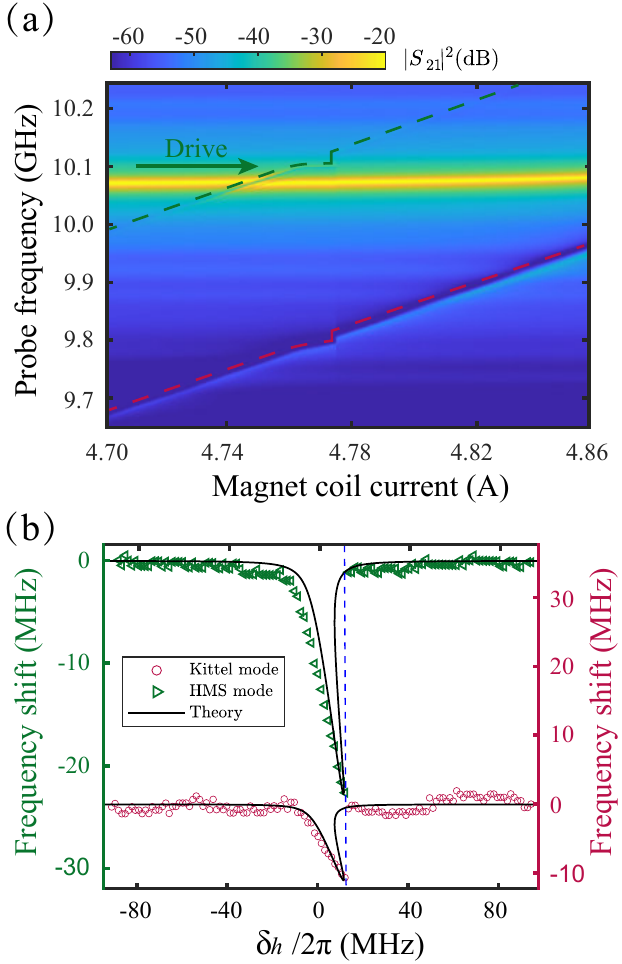}
		\caption{Drive on the HMS mode. (a) The transmission spectra are measured versus magnet coil current, while a drive field of 10.1 GHz and 25 dBm is applied on the HMS mode. The HMS mode has a frequency shift $\Delta_{h,s}$ due to the self-Kerr effect, and the Kittel mode has a smaller frequency shift $\Delta_{k,c}$ caused by the cross-Kerr effect. (b) The frequency shifts of the Kittel mode and HMS mode are extracted from (a), and plotted versus $\delta_{h} = \omega_{h} -\omega_d $. The experimental results are fitted by Eq.~(\ref{eqution3}) and $\Delta _{k,c}=\frac{K_{cross}}{K_{h,s}}\Delta_{h,s}$, which gives the ratio $K_{cross}/K_{h,s}=0.49$. }
		\label{Figure3}
	\end{figure}

	\subsection{Drive the HMS mode}
	The cross-Kerr effect implies that the HMS mode excitation will also induce a frequency shift of the Kittel mode. Next, we apply the drive field (25 dBm) at a fixed frequency of 10.1 GHz on the HMS mode and sweep the magnetic field. The measured transmission mapping is shown in Fig.~\ref{Figure3}(a). As expected, the HMS mode has a frequency shift due to the self-Kerr effect, and simultaneously the cross-Kerr effect shifts the Kittel mode frequency. We extract the frequency shifts of these two modes versus the drive field detuning $\delta_{h}=\omega_{h}-\omega_{d}$ as shown in Fig.\ref{Figure3}(b). Similarly, the self-Kerr effect induced frequency shift $\Delta _{h,s}$ of the HMS mode obeys the following equation:
	\begin{equation}
		\left[ {{{\left( {{\Delta _{h,s}} + {\delta _h}} \right)}^2} + {(\gamma _h/2)}^2} \right]{\Delta _{h,s}} - {c_h}{P_d} = 0,
		\label{eqution3}
	\end{equation}
	where ${\gamma _{h}/2\pi}=5.0$ MHz is the HMS mode linewidth, $c_h$ is the coefficient of drive efficiency on the HMS mode. The frequency shift $\Delta_{k,c}$ of the Kittel mode induced by the cross-Kerr effect can be further fitted by ${\Delta _{k,c}}=\frac{K_{cross}}{K_{h,s}}\Delta_{h,s}$. The fitting curves are shown as black lines in Fig.~\ref{Figure3}(b). We get $K_{cross}/K_{h,s} = 0.50$.
	
	From the above two sets of experiments, we get $K_{h,s}/K_{k,s}=5$, which shows that the HMS mode's self-Kerr nonlinearity is five times larger than that of the Kittle mode. In any case, the cross-Kerr effect plays a vital role in the nonlinear process, and it acts as a ruler telling us the ratio between the nonlinear coefficients of different spin wave modes. More specifically, by comparing the amounts of the frequency shifts observed in the two sets of experiments where the drive powers are both 25 dBm, we can conclude that pumping the HMS mode is more challenging than pumping the Kittel mode with the same loop antenna, which may be caused by the small dipole moment of the HMS mode.
	
	\begin{figure}[!t]
		\centering
		\includegraphics[width=0.43\textwidth]{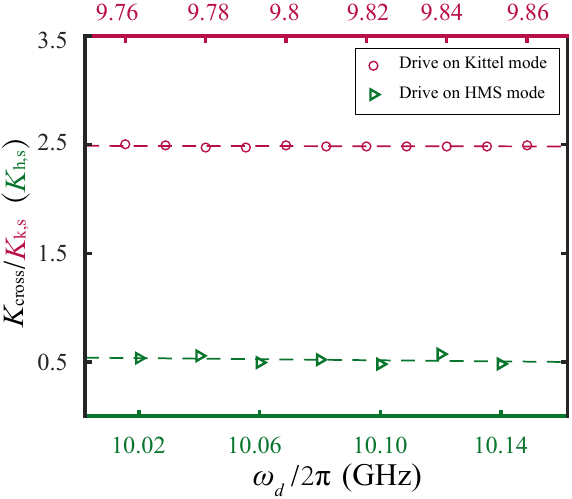}
		\caption{The ratios between the cross-Kerr and the two self-Kerr coefficients at different drive field frequencies. The ratios are found to be nearly constant. The mean value of $K_{cross}/K_{k,s}$ is 2.49 and the mean value of $K_{cross}/K_{h,s}$ is 0.53.}
		\label{Figure4}
	\end{figure}
	
	\subsection{Stable feature under various drive frequencies}
	Finally, we carry out a series of experiments where the drive field frequency is changed. We measure the ratios of the cross-Kerr to the two self-Kerr coefficients. As shown in Fig.\ref{Figure4}, the ratios keep consistent at different drive field frequencies, which imply that the observed nonlinearity is the intrinsic property of this ferrimagnetic crystal. This stable feature provides more guarantees for the information conversion and storage among different modes in a single YIG sphere.

	\section{CONCLUSION}
	The cross-Kerr effect is an important nonlinearity, which describes the effect of excitations in one mode on another mode's resonant frequency. In this work, we investigate the cross-Kerr effect in a ferrimagnetic spin ensemble, in which one Kittel mode and one higher-order magnetostatic mode co-exist. The cross-Kerr interaction between them plus their individual self-Kerr effects give rise to fascinating phenomena. In the experiment, we drive the Kittel model and the higher-order magnetostatic mode, respectively, and obtain the ratios between the cross-Kerr coefficient and their self-Kerr coefficients. Our experiment opens up a new path to study nonlinear effects in magnetic materials and spin ensembles. The mutual interaction between different magnetostatic modes in a single YIG sample can also provide new degrees of freedom for cavity spintronics and cavity magnonics~\cite{Roadmap}. We anticipate that this finding may stimulate more designs and applications of cavity magnonics.

	\begin{acknowledgments}
		The authors thank Y. H. Li for helpful discussions. This work is supported by the National Natural Science Foundation of China (Grants No. 11934010, No. U1801661, and No. 12174329), Zhejiang Province Program for Science and Technology (Grant No. 2020C01019), the Fundamental Research Funds for the Central Universities (No. 2021FZZX001-02), and the China Postdoctoral Science Foundation (No. 2019M660137).
	\end{acknowledgments}
	
	\appendix
	\renewcommand{\appendixname}{APPENDIX}
	\begin{figure*}[htb]
		\centering
		\includegraphics[width=0.8\textwidth]{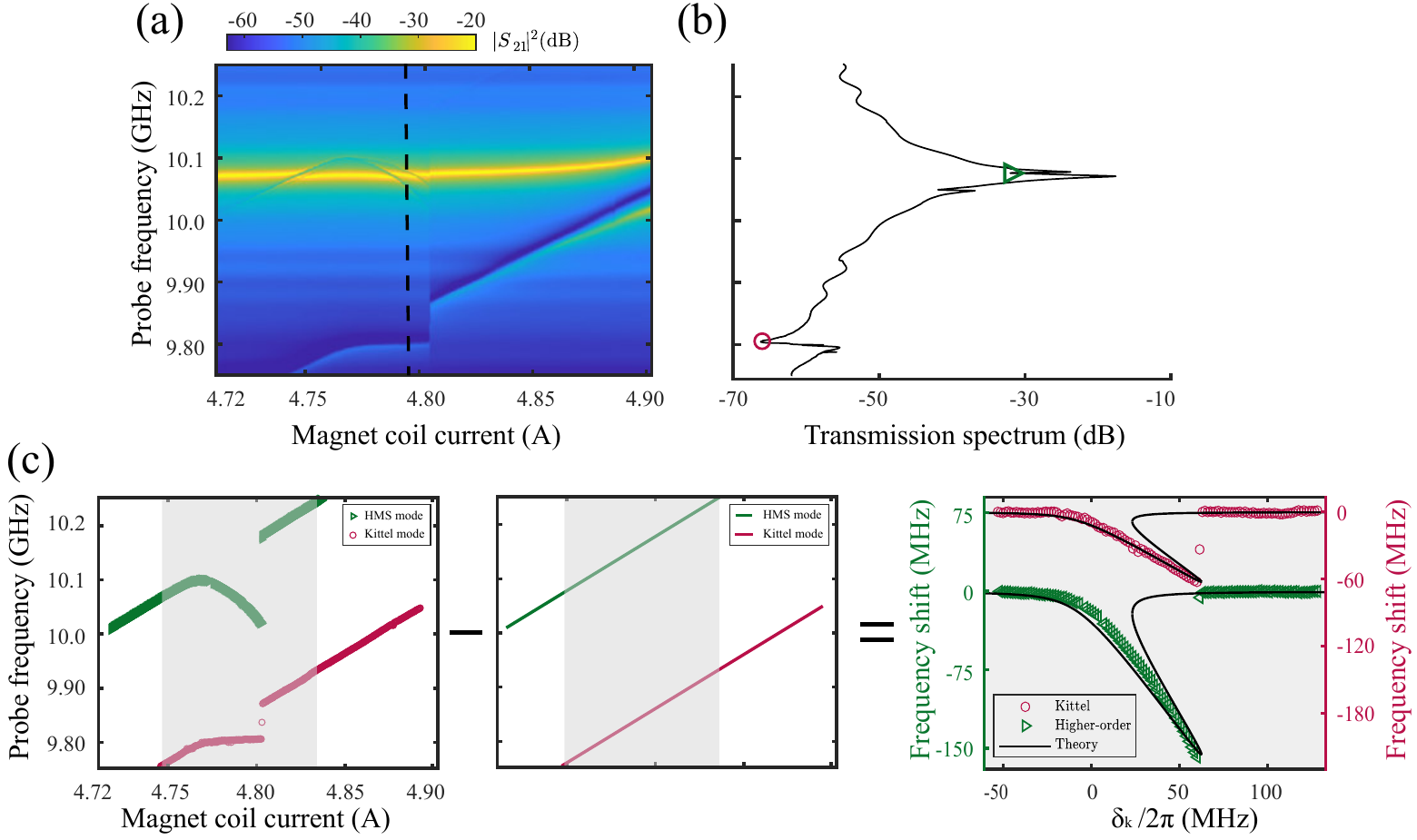}
		\caption{(a) Original transmission spectrum mapping. (b) The transmission spectrum $|S_{21}|^2 $ correspond to the black dashed line labelled in (a) at a certain bias magnetic field. (c) The Kerr effects induced frequency shifts are extracted. We subtract the linear frequency shift caused by the external bias field from all data points. The horizontal axis is changed to ${\delta _k} = {\omega _k}- {\omega _d}$ for fitting with Eq.~(\ref{eqution2}).}
		\label{FIG.A1}
	\end{figure*}
	\section{INTERACTION BETWEEN THE KITTEL MODE AND HMS MODE}
	First, we derive the Hamiltonian of the magnon modes without including the cross-Kerr effect. Under the bias magnetic field $B_0$, the Hamiltonian of the Kittel mode is written as (HMS mode has a similar form)~\cite{Y.P.Wang2016}
	\begin{equation}
		{H_k} =- \int\limits_{{V}} {{\bf{M}}_K}  \cdot {{\bf{B}}_0}d\tau  - \frac{{{\mu _0}}}{2}\int\limits_{{V}} {{\bf{M}}_K \cdot {{\bf{H}}_{\rm{an}}}d\tau } ,
		\label{A2}
	\end{equation}
	where the first term represents the Zeeman energy and the second term is the magnetocrystalline anisotropy energy. ${\bf{M}}_K=\left( {{M_{K,x}},{M_{K,y}},{M_{K,z}}} \right)$ is the sub-magnetization corresponding to the Kittel mode. $V$ is the volume of the YIG sphere, $\mu_{0}$ is the vacuum permeability, and $\mathbf{H}_{\rm{an}}$ is the anisotropic field due to the magnetocrystalline anisotropy in the YIG crystal.
	
	We adopt the direction of the bias magnetic $\mathbf{B}_{0}$ as the $z$ direction  ($\mathbf{B}_{0}=B_{0}\mathbf{e}_{z}$). When the [110] crystal axis of the YIG sphere is aligned along the bias magnetic field, the anisotropic field is given by~\cite{stancil2009}
	\begin{equation}\label{Han}
		\mathbf{H}_{\rm{an}}=\frac{3K_{\rm{an}}M_{x}}{\mu_{0}M^{2}}\mathbf{e}_{x}+\frac{9K_{\rm{an}}M_{y}}{4\mu_{0}M^{2}}\mathbf{e}_{y}+\frac{K_{\rm{an}}M_{z}}{\mu_{0}M^{2}}\mathbf{e}_{z},
	\end{equation}
	where $K_{\rm{an}}$ is the first-order magnetocrystalline anisotropy constant. Then, the Hamiltonian of the Kittel mode has the form
	\begin{eqnarray}\label{e0002}
		H_k&=&-B_{0}M_{K,z}V-\frac{3K_{\rm{an}}M_{K,x}}{2M_{K}^{2}}\mathbf{e}_{x}\nonumber\\&-&\frac{9K_{\rm{an}}M_{K,y}}{8M_{K}^{2}}\mathbf{e}_{y}-\frac{K_{\rm{an}}M_{K,z}}{2M_{K}^{2}}\mathbf{e}_{z}.
	\end{eqnarray}
	
	By using the macrospin operator $\mathbf{S}=\mathbf{M}V/\hbar\gamma\equiv(S_{x},S_{y},S_{z})$, where $\gamma$ is the electronic gyromagnetic ratio~\cite{Soykal2010}, the Hamiltonian $H_{k}$ can be written as
	\begin{eqnarray}\label{e0003}
		H_k/\hbar&=&-\gamma B_{0}S_{K,z}-\frac{3\hbar K_{\rm{an}}\gamma^{2}}{2M_{K}^{2}V}S_{K,x}^{2}\nonumber\\&-&\frac{9\hbar K_{\rm{an}}\gamma^{2}}{8M_{K}^{2}V}S_{K,y}^{2}-\frac{\hbar K_{\rm{an}}\gamma^{2}}{2M_{K}^{2}V}S_{K,z}.
	\end{eqnarray}

	We harness the Holstein-Primakoff transformation~\cite{Holstein1940}: $S_{K}^{+}=\sqrt{2 S_{K}-b^{\dagger} b}b$, $S_{K}^{-}=b^{\dagger} \sqrt{2 S_{K}-b^{\dagger} b}$, and $S_{K,z}=S_{K}-b^{\dag}b$, where $S_{K}$ is the total spin number, and $S_{K}^{\pm}=S_{K,x}\pm iS_{K,y}$. The excited magnon number is small compared with the total spin
	number $S_{K}$, and the higher-order terms can be neglected in the Holstein-Primakoff transformation. In this case, we have $S^{+}\approx \sqrt{2S}b$ and $S^{-}\approx b^{\dag}\sqrt{2S}$. Under the rotating-wave approximation, we finally obtain the Hamiltonian
	\begin{equation}
		{H_k}/\hbar = {\omega _{k'}}{b^\dag }b + {K_{k,s}}{b^\dag }b{b^\dag }b,
		\label{A3}
	\end{equation}
	where $${K_{k,s}}=13\hbar K_{\rm{an}}\gamma^{2}/(16M_{K}^{2}V),$$
	and the value of $K_{\rm{an}}$ can be found in Ref.~\cite{stancil2009}.
	
	For the interaction Hamiltonian between the Kittel mode and HMS mode, we consider these two modes as two sets of submagnetization, and the interaction between them can be written as
	\begin{equation}
		\begin{split}
			H_{k,h}&= \beta \int\limits_{V}{\textbf{M}_{K} \cdot {\textbf{M}}_H} d\tau \\
			&= \frac{\beta\hbar^{2} \gamma ^2}{V}\textbf{S}_K \cdot \textbf{S}_H\\
			&= \frac{\beta\hbar^{2} \gamma ^2}V(S_{K,x}S_{H,x} + S_{K,y}S_{H,y} + S_{K,z}S_{H,z}),
		\end{split}
		\label{A6}
	\end{equation}
	where $\beta$ is a coefficient that measures the mode overlapping between the Kittel mode and HMS mode. This mode overlap originates from the partial local spins shared by the Kittel mode and other spin wave modes~\cite{A.Gloppe2019,stancil2009}. Under the Holstein-Primakoff transformation and rotating-wave approximation, we can get
	\begin{equation}
		\begin{split}
			{H_{k,h}}& = \frac{\beta\hbar^{2} {\gamma ^2}}{V} [-S_H{b^\dag}b-S_K{c^\dag }c \\
			& +{b^\dag}b{c^\dag}c + \sqrt{{S_K}{S_H}} ({b^\dag}c+b{c^\dag})].\\				
		\end{split}
		\label{A7}
	\end{equation}
	The first and second terms produce constant frequency shifts in the Kittel mode and HMS mode, respectively. The third term refers to the coherent coupling between the Kittel mode and HMS mode. In our experiment, the detuning between the Kittel mode and HMS mode is sufficiently large, and the system is in the dispersive regime. If we tune the Kittel mode close to the HMS mode, we can predict level repulsion. A related effect has been observed in a previous study~\cite{Paul}.
	
	The total Hamiltonian of the cavity magnonic system becomes
	\begin{equation}\label{A9}
		\begin{split}
			H/\hbar &= \omega _{c}a^\dag a + \omega _{k}b^\dag b+ \omega_{h}c^\dag c\\
			&+K_{k,s}b^\dag bb^\dag b + K_{h,s}c^\dag cc^\dag c+K_{cross}b^\dag bc^\dag c \\
			&+g_{k}\left(a^\dag b + b^\dag a \right)+g_{h}\left( a^\dag c + c^\dag a \right)+g_{k,h}\left( b^\dag c + c^\dag b \right)\\
			&+\Omega _k(b^\dag e^{ - i{\omega _d}t} + b e^{i{\omega _d}t})\\
			&+\Omega _h(c^\dag e^{ - i{\omega _d}t} + c e^{i{\omega _d}t}),
		\end{split}
	\end{equation}
	where ${\omega _k}={\omega _{k'}} - {\beta\hbar {\gamma ^2}{S_H}}/{V},
	~{\omega _h} = {\omega _{h'}} - {\beta \hbar{\gamma ^2}{S_K}}/{V},
	~{K_{cross}}={\beta\hbar {\gamma ^2}}/{V},$ and $g_{k,h}=\beta\hbar {\gamma ^2}\sqrt{{S_K}{S_H}}/{V}$.
	
	\section{DATA PROCESSING METHOD}
	The transmission spectrum mapping consists of a sequence of transmission spectra measured at different bias magnetic fields, as shown in Fig.~\ref{FIG.A1}(a). The individual transmission spectrum is depicted in Fig.\ref{FIG.A1}(b). The anti-resonances correspond to various modes; for example, the green triangle dot represents the HMS mode, and the circle red dot denotes the Kittel mode. We extract the data points from the sequence of transmission spectra and plot them in the left panel of Fig.\ref{FIG.A1}(c). To obtain a plot of frequency shift versus driving detuning that is easy to fit with our theoretical model, we subtract the linear component that increases with the external bias field from all data points. Then, the remaining component is the contribution of frequency shift caused by the driving field and Kerr effects. The middle panel of Fig.~\ref{FIG.A1}(c) depicts the linear frequency shift caused by the increased bias magnetic field. In order to fit the experimental data with Eq.~(\ref{eqution2}), we replace the horizontal axis with drive detuning ${\delta _k} = {\omega _k}- {\omega _d} $.

\end{document}